\documentclass[aps,preprint,superscriptaddress,groupedaddress]{revtex4}  
\usepackage{color,graphicx}  
\usepackage{dcolumn}   
\usepackage{bm}        
\usepackage{amssymb}   
\usepackage{amsmath}
\DeclareGraphicsExtensions{.pdf,.pgn,.jpg, .eps}
\usepackage{subfig}
\usepackage{hyperref}
\hypersetup{
    colorlinks,
    citecolor=black,
    filecolor=black,
    linkcolor=black,
    urlcolor=black
}

\begin{document}

\widetext

\title{Effective spreading from multiple leaders identified by percolation in social networks}


\date{\today}

\author{Shenggong Ji}
\affiliation{School of Mathematics, Southwest  Jiaotong University, Chengdu 610031, China}

\author{Linyuan L\"u}
\affiliation{Alibaba Research Center for Complexity Sciences, Hangzhou Normal University, Hangzhou 310036, China}

\author{Chi Ho Yeung}
\affiliation{Department of Science and Environmental Studies, The Hong Kong Institute of Education, Hong Kong}

\author{Yanqing Hu}
\affiliation{School of Mathematics, Southwest  Jiaotong University, Chengdu 610031, China}

\begin{abstract}
Social networks constitute a new platform for information propagation, but its success is crucially dependent on the choice of spreaders who initiate the spreading of information. In this paper, we remove edges in a network at random and the network segments into isolated clusters. The most important nodes in each cluster then form a group of influential spreaders, such that news propagating from them would lead to an extensive coverage and minimal redundancy. The method well utilizes the similarities between the pre-percolated state and the coverage of information propagation in each social cluster to obtain a set of distributed and coordinated spreaders. Our tests on the Facebook networks show that this method outperforms conventional methods based on centrality. The suggested way of identifying influential spreaders thus sheds light on a new paradigm of information propagation on social networks.

\end{abstract}

\maketitle

\section{Introduction}

The development of social network has a great impact on our lifestyle, from making friends to dating, from working to shopping. They become more essential as we are increasing our dependence on them to gather information. Compared with search engines which are based on isolated queries, collecting information through leveraging the individual specialties in social networks lead us to useful websites from experts in disparate fields, and thus increase both the quality and the diversity of the acquired information. Thus, by the same token, influential individuals can also be used to spread information. The key to the success is to identify the most influential spreaders in the network. Nevertheless, it is difficult as there are usually just a few users capable of propagating a news to a large number of users \cite{Albert}. For example, while socially significant users are rare in the tweeter network, their messages and blogs can spread quickly throughout the whole network \cite{Wen2010,hu2014conditions}.

Although social networks are powerful for propagating information, their application for this purpose is limited, partially because a way to identify the optimal spreaders is absent. Nevertheless, simple methods have been proposed. For instance, ``degree centrality" suggests that nodes with higher degree are more influential than the others \cite{Wasserman}. On the other hand, the location of a node in a network and the influence of its neighbors are also considered important. For instance, a node with a small number of highly influential neighbors located at the center of the network may be more influential than a node having a larger number of less influential neighbors. Kitsak et al. \cite{Kitsak} thus proposed a coarse-grained method to use the $k$-core decomposition to quantify the influence of a node, based on the assumption that news initiated at nodes in higher shells are likely to spread more extensively. Some distance-based global metrics such as betweenness \cite{Freeman1977} and closeness \cite{Sabidussi} are suggested which can lead to extensive propagation, but due to the high computational complexity, they are not practical for large-scale social networks. Other centralities such as LocalRank were also suggested \cite{ChenDB}.

The above simple but sub-optimal protocols have been applied to social media such as QQ , BBS and Blog to find the key spreaders who can trigger the ``tipping point" in social marketing to promote commercial products. Specifically, if one can convince a set of influential users to adopt a new product, one may induce a large cascade of purchases as these initial buyers propagate their compliment of the product along the network. Unlike the forementioned methods which identify a set of independent spreaders according to their centralities, our goal is to find a set of coordinated individuals such that their combined impact is greatest, leading to much more extensive propagation of information. Nevertheless, identifying the optimal group of spreaders is indeed a computationally difficult task \cite{Kleinberg}.

In this paper, we utilize the similarities \cite{PastorSatorras,Newman2002} between percolation and information propagation to identify a group of influential spreaders. By removing edges at random until percolation ceases, individual isolated clusters are formed. Due to the correspondence between percolation and information transmission, the emergence of such clusters imply that news can be effectively propagated within the clusters but not across the clusters. Initiating a news on the most influential user in each cluster is thus an effective way to distribute the news within the cluster. Since such process is static and requires much less computation power than the dynamics spreading of news, a lot of percolated states can be generated to give a more accurate result on the segmentation of social clusters as well as their corresponding influential spreaders.

By testing our protocol on Facebook and Enron email network, we show that in addition to a lower computational efficiency, our protocol outperforms other simple heuristics based on local and global centrality in terms of propagation coverage and coverage redundancy of the selected spreaders. This is consistent with the old saying that the power of a typical group exceeds that of a single most competent individual. Moreover, we find that the average degree of the users selected by our method is lower, which implies a lower cost in identifying the spreaders when compared to the other methods. We also identify the different characteristics of spreaders who are most effective to promote niche or popular items in order to maximize the coverage. All these results lead to insights into the design of viral marketing strategies and a new paradigm for information propagation.

\section{Results}

Spreading dynamics with the involvement of human can be mainly classified into two classes: one is the spreading of infectious diseases which requires physical contacts, and the other is the spreading of information, including opinions and rumors where physical contacts are not required \cite{LuNJP2011}. Due to the similarity between epidemic and information spreading, well-established models of epidemic models are widely used to describe the propagation of information \cite{Sudbury1985,Zanette2001,Zanette2002,LiuZ2003,Moreno2004}.

In particular, the susceptible-infected-recovered (SIR) model is one of the representatives. Specifically, a susceptible person (S) in the model is analogous to an individual who is not aware of the information. An infected person (I) is analogous to an individual who is aware of the information and will pass it to his/her neighbors. A recovered person (R) is analogous to an individual who loses his/her interest and will never pass the information again. Newman \cite{Newman2002} studied in detail the relation between the static properties of the SIR model and bond percolation phenomenon on networks and remarked that the SIR model with transmissibility $p$ is equivalent to a bond percolation model with bond occupation probability $p$ on the network (see Method and Materials 4.5).

Our method is then devised in relation to the bond percolation model as follows. Given an undirected network $G(V, E)$ where $V$ represents the set of nodes (i.e., users in social networks) and $E$ represents the set of edges (i.e., connection in terms of communication, friendship or other kinds of interactions), all edges are first removed and each individual edge is then recovered with a probability $p$, i.e. all links are removed when $p=0$. As $p$ increases from $p=0$, more links are recovered and clusters start to form and merge with each other. We will call this state the \emph{pre-percolated state}. For a network containing $N$ node, a giant component of size $O(N)$ emerges only when $p$ is larger than a critical threshold $p=p_c$, which is called percolation. In the context of information propagation, since an edge between two nodes appears with a probability $p$, the value $p$ can be considered as the transmissibility of an information from one node to another.

To find the influential group, we have to find the $W$ most influential spreaders with a given value of $p$. Assume that there are $m$ percolation clusters after one realization of link recovery, and denote by $S_i$ the size of cluster $i$, $i=1,2,\cdots,m$. We introduce a tunable parameter $L$, which is usually equal to or larger than $W$. If $L\leq m$, we choose the top-$L$ largest clusters and assign one score to the largest degree node in each cluster. If there are many nodes with the largest degree, we assign the score to a random one among them. If $m<L\leq 2m$, we first choose the top-$L$ node in each cluster, and the rest $L-m$ nodes are chosen to be those with the second largest degree respectively from the top-$(L-m)$ largest clusters. If $L>2m$, we will choose the next largest degree nodes in each cluster following the same selection rules. After $t$ times of different trials of link recovery, all nodes are ranked according to their scores in a descending order and those $W$ nodes with the highest scores are suggested to be the set of initial spreaders. For the sake of simplicity, we set $L=W$ and have tested and found that the results are not sensitive to $L$. The dependence of $L$ are shown in the Supplementary Information (SI) Sec. SV Fig. S6.

In other words, our suggested method draws analogy with percolation to identify individual social clusters in the network where news can be effectively propagated within the clusters but not across the clusters. These isolated clusters in the pre-percolated state thus have a direct correspondence to the propagation coverage when one spreads the news from an initial spreader in each of the clusters. Our rationale is different from most other methods which usually identify a group of influential spreaders for the network as a whole. In addition, such a set of well distributed spreaders also enjoy a reduced redundancy when compared to a set of un-coordinated spreaders. These differences make our method unique compared to the other methods.

\subsection{Spreadability and coverage redundancy}

To quantify the performance of our method, we examine the spreadability, i.e. the propagation coverage of a news from a set of $k$ selected spreaders, by our method as well as other methods. We will use the SIR model to mimic the spreading of news, and the spreadability is defined as the ratio of recovered nodes to the total number of nodes (i.e., the size of outbreak to $N$). We remark that the transmissibility $p$ adopted in the SIR model is the same as the probability $p$ used to recover edges to identify the clusters in the percolated states. As a result, for a single spreader, the ultimate size of the SIR outbreak triggered by this spreader is precisely the size of the percolation cluster that it belongs to. Likewise, the ultimate size of the SIR outbreak triggered by a group of spreaders in distinct percolation clusters is the sum of the size of the clusters that these nodes belong to. For example, if we measure the coverage of three selected nodes on the network with $N$ nodes, and if the first two nodes belong to the cluster $S_1$, and the third one is in the cluster $S_2$. For each of the single nodes, the coverage of node 1,2 and 3 are respectively $S_1/N$, $S_1/N$ and $S_2/N$, while for the whole group, the spreadability of the three nodes is $(S_1+S_2)/N$.

We first apply our method on the Facebook network with 59691 nodes. Figure 1a shows the coverage obtained from 4000 initial spreaders chosen by our percolation method, compared with a set of 4000 spreaders identified by three other methods, namely the degree centrality, the $k$-shell decomposition and the betweeness centrality (see \textbf{Methods and Materials} for the definition of each of these methods; comparisons with other centrality measures can be found in Sec. SIII Fig. S2 of the SI). Percolation method yields the highest spreadability for an arbitrary transmissibility $p$ . Figure 1b shows the degree distribution of the 4000 spreaders identified by the percolation method.

When $p<p_c = 0.01$,the percolation method yields isolated clusters \cite{Newman2001} (see Fig. 1d) of similar size, and since the set of the selected spreaders come from different clusters, and a wide range of degree is found among the spreaders (see Fig. 1b). In this case, the percolation method is more likely to choose high-degree nodes (see Fig. S5b in the SI, where the red stars represent the degree distribution of the 4000 selected nodes when $p=0.008$). When $p>p_c$, the distribution will become narrower as $p$ increases (see the blue squares in Fig. S5b of the SI). In this case, the percolation method prefers low-degree spreaders. The average original degree  (i.e. degree in the original network before edge removal) of the 4000 spreaders selected by the percolation method when $p<p_c$ is higher than that of the nodes selected when $p>p_c$. This implies that if we want to promote and advertise a new niche product which is difficult to get accepted, one can draw analogy with the case of small transmissibility $p$ where high-degree initial spreaders are preferred. On the other hand, for popular items which are easy to be accepted, one can draw analogy with the case of large $p$ and low-degree initial spreaders are preferred.

We then examine the cost of identifying the initial spreaders. By assuming that the direct influence of a user is equal to the number of its nearest neighbors (i.e., its degree),  while the difficulties of finding a user with degree $k$ is proportional to $1/p(k)$, the cost to find a spreader $i$ is assumed to be $k_i/p(k_i)$. Figure 1c shows the dependence of the average cost to find the 4000 spreaders under the parameter $p$, i.e. $\frac{1}{W}\sum_{i=1}^{W} \frac{k_i}{p(k_i)}$. The cost decreases abruptly at the critical point $p_c$, indicating a phenomenon resembling phase transition. It means that when $p$ increases just beyond $p_c$, the cost can be reduced substantially.

Besides the spreadability and the cost, we also examine the redundancy in coverage which quantifies the efficiency of the propagation. Specifically, the redundancy of a node $i$ is defined as the number of initial spreaders who has the potential to infect node $i$. A method is inefficient if the initial chosen spreaders pass the same information to the same group more than once. Averaging the redundancy over all the infected nodes, we obtain the redundancy of the set of initial spreaders. Figure 1e compares the spreading redundancy of our method with the three other methods (comparisons with other centrality measures can be found in Sec. SIII Fig. S3 of the SI). Highest redundancy is found in the methods of $k$-shell and degree centrality, followed by betweenness centrality. Our percolation method has the lowest redundancy among the four methods, since the spreaders identified by this method are usually located in different regions of the original network. We also checked the Enron e-mail network and similar results with Facebook network are obtained (see Supplementary Sec.II figure S1).

To further examine the spreadability, we applied the four methods to identify four initial spreaders on a generated network with four clear communities. As shown in Fig. 2, the four spreaders identified by the percolation method are very likely to be found in different communities, with one spreader in each community. For the other methods, there are high probabilities that all or some of the initial spreaders are in the same communities. These results are easy to understand as our method relies on the segmentation of the network into isolated clusters to identify the spreaders. In the present case, the network is likely to separate into the four communities and thus one spreader is found in each community.

\begin{figure}
\begin{center}
\includegraphics[width=16cm]{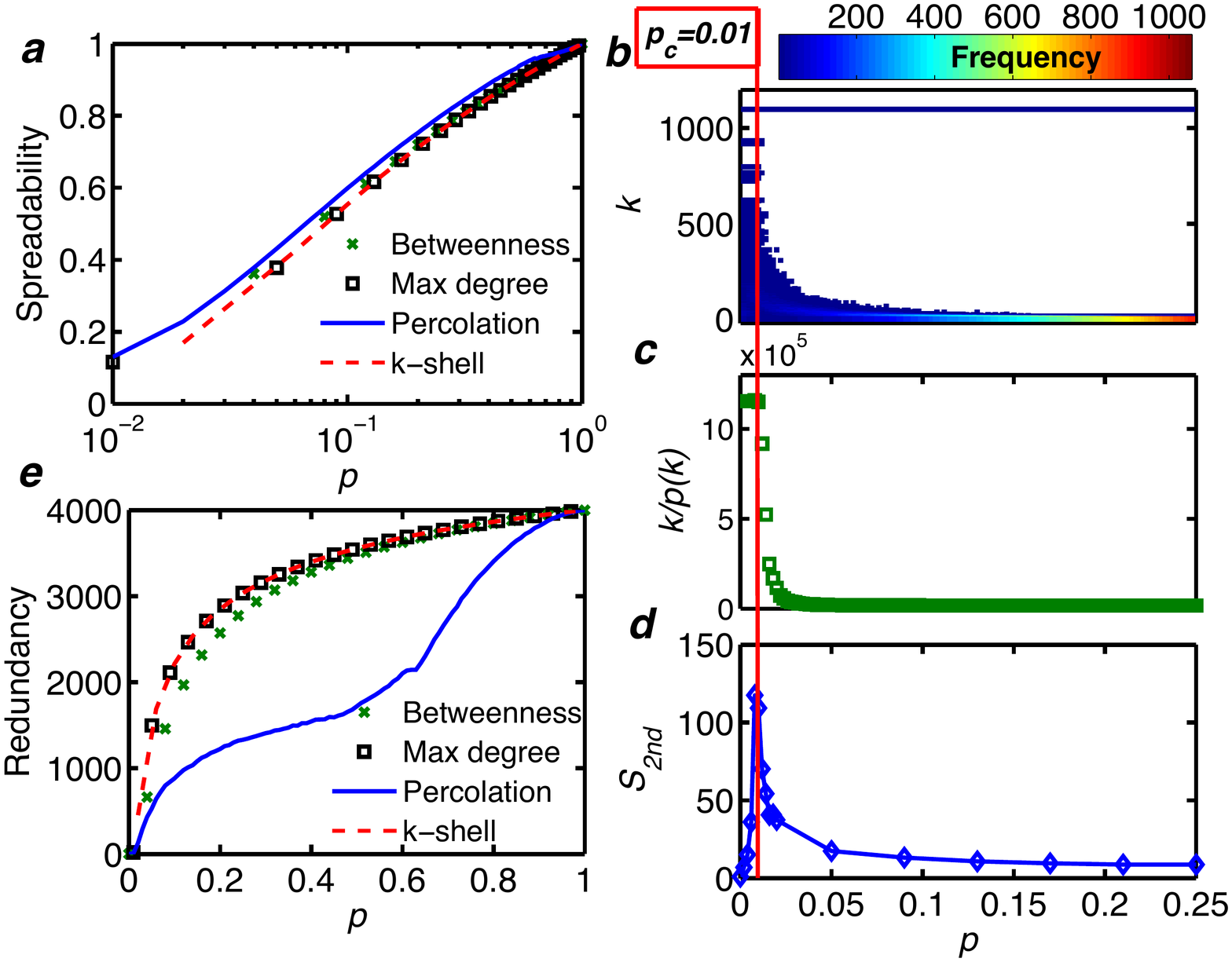}
\caption{The performance of the percolation method on the Facebook network. (a) The spreadability of the 4000 spreaders selected by the percolation method (blue solid line), maximum degree (squares), $k$-shell (red dash line) and betweeness centrality (dots). (b) The degree distribution s of the 4000 initial spreaders selected by the percolation method given different transmissibility $p$. The color indicates the frequency, blue to red corresponds to low to high. (c) The cost to identify such a group of spreaders. (d) The size of the second giant component, $S_{2nd}$, after percolation with different transmissibility $p$. The largest value of $S_{2nd}$ \cite{Parshani2011} is obtained at the critical point $p_c=0.01$. (e) The coverage redundancy of the four methods.}\label{example}
\end{center}
\end{figure}


\begin{table}
\caption{The percentage of the different distribution of the four spreaders in a model network with four communities. For instance, the four spreaders identified by our methods are found in four different communities in 93.9$\%$ of the realizations. The procedure to construct the networks is shown in Method and Materials. The results are obtained based on 1000 realizations. The critical point of the network is $p_c \approx 0.2687\pm 0.0152$. We have set $p=0.28$ in our percolation method.}
\centering
\begin{tabular}{|c|c|c|c|c|c|}
\hline
Type & \includegraphics[width=2cm]{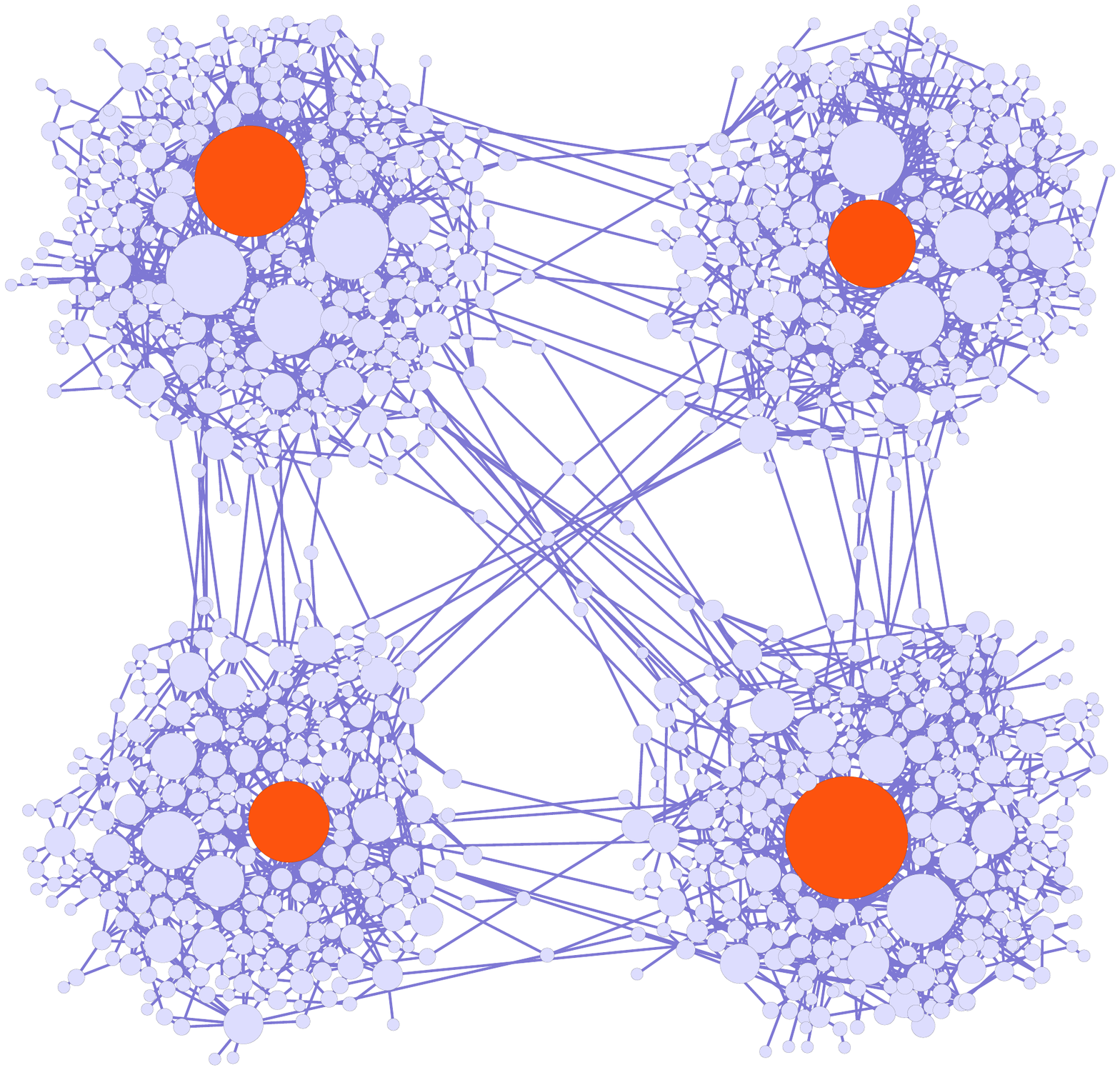} &\includegraphics[width=2cm]{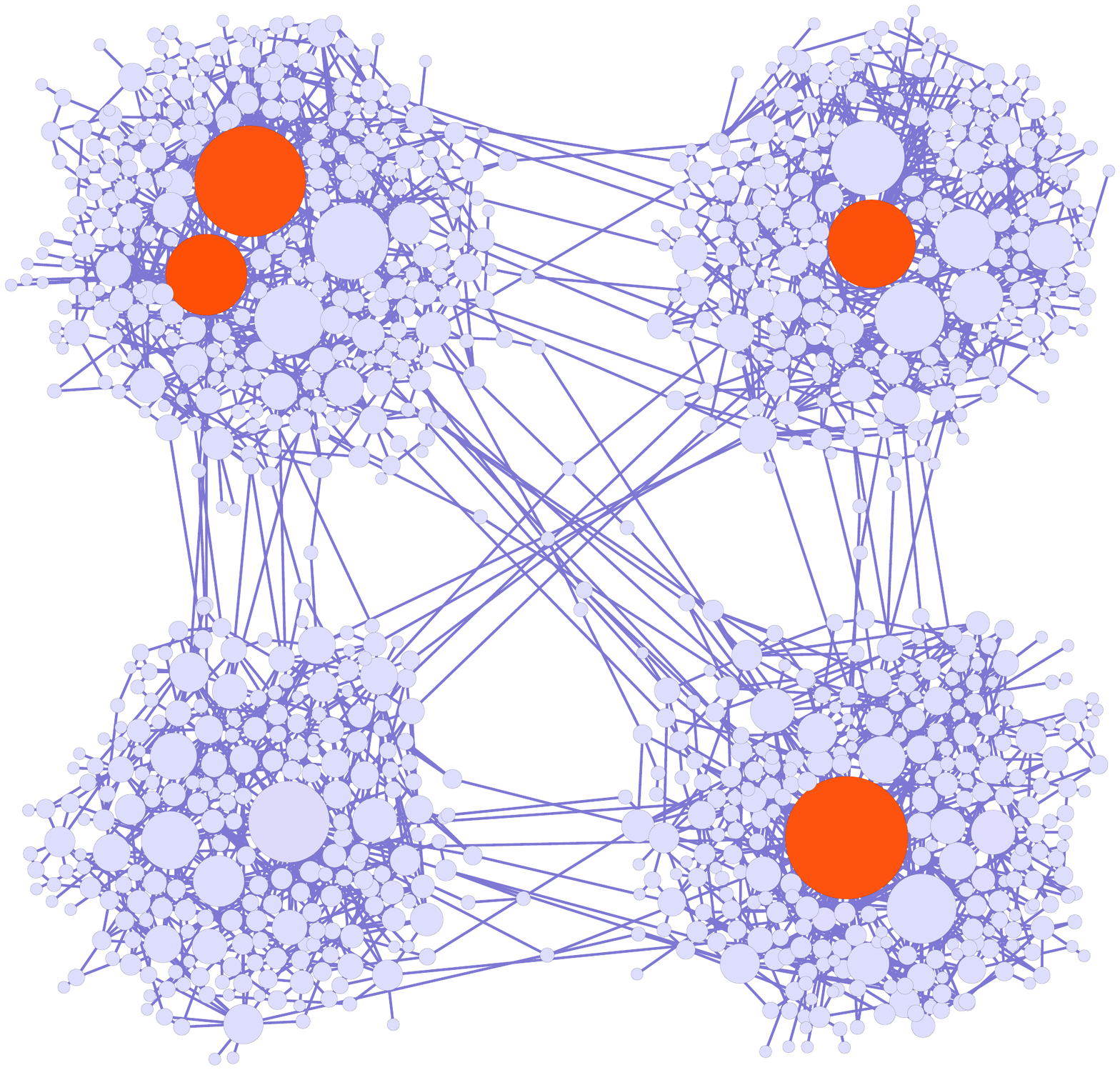}  &\includegraphics[width=2cm]{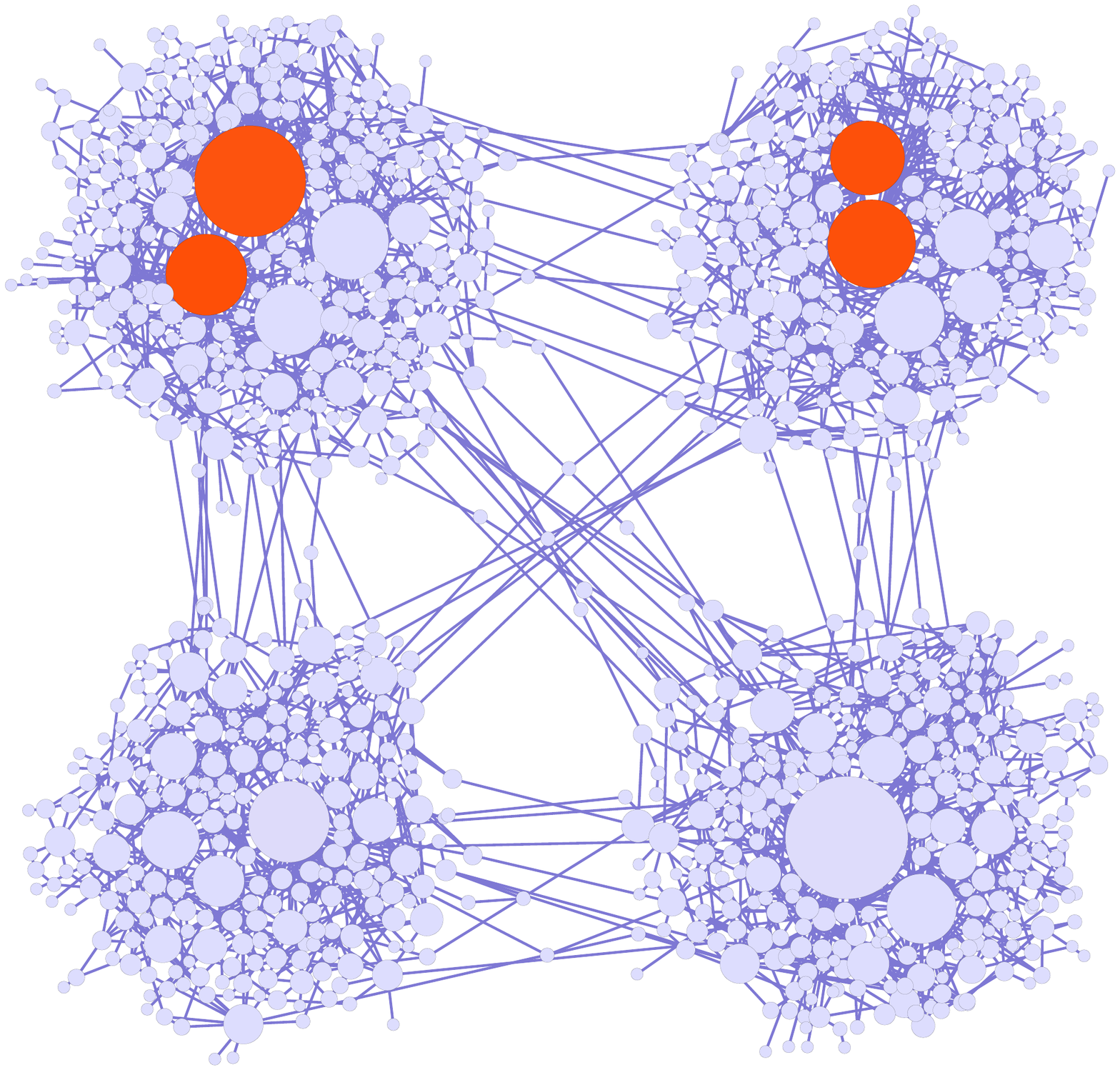} &\includegraphics[width=2cm]{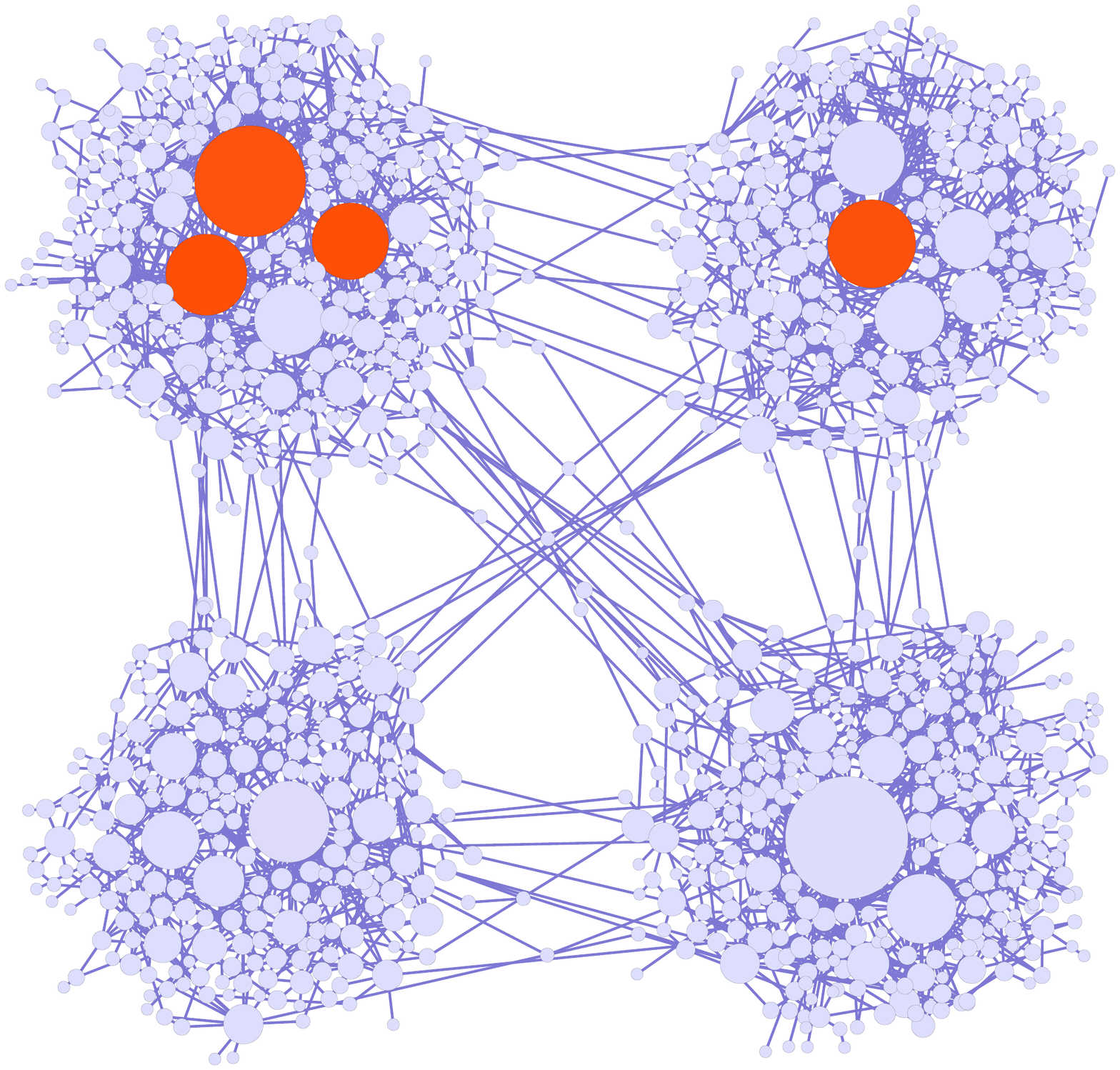}  &\includegraphics[width=2cm]{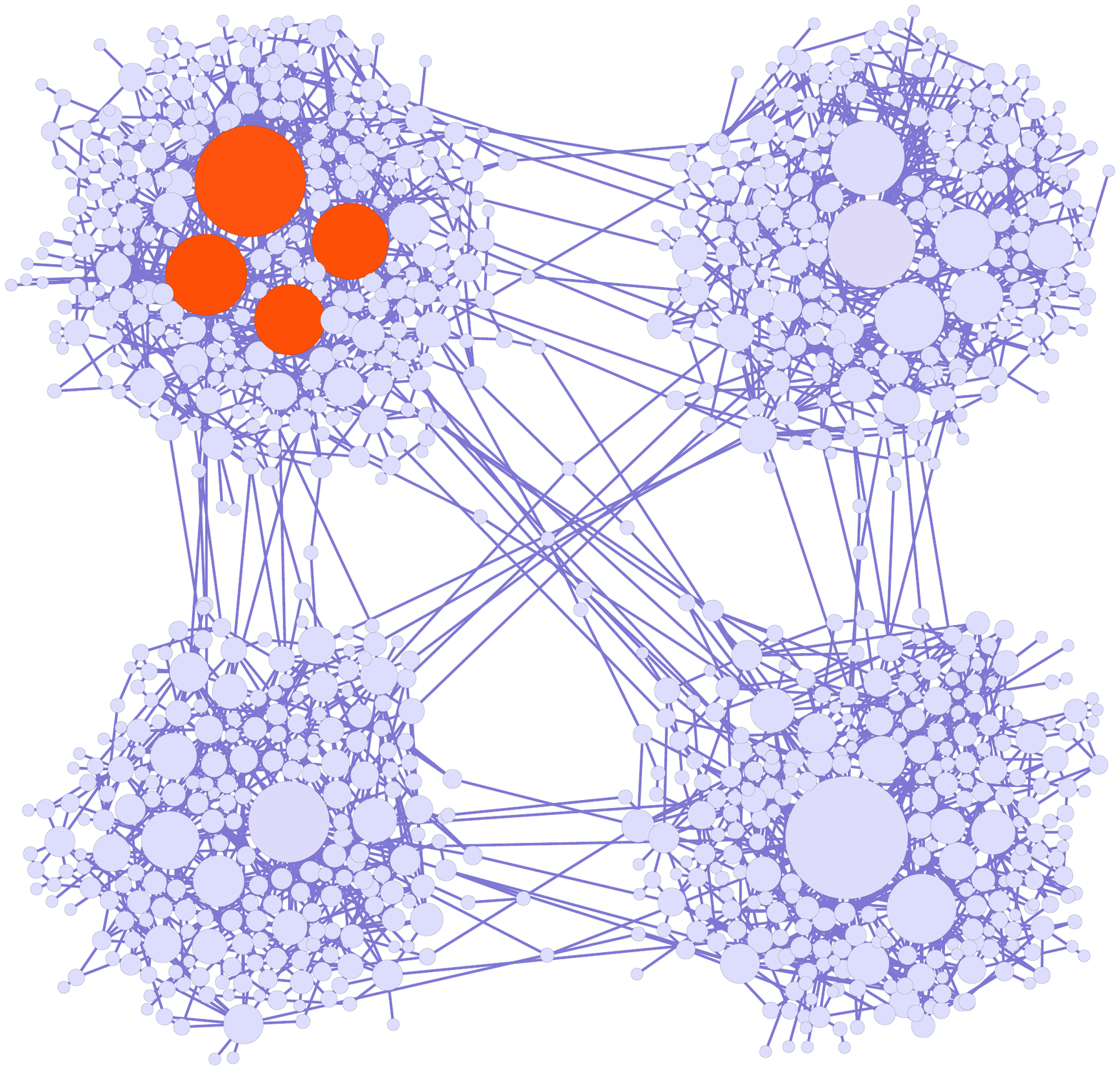} \\
\hline
Percolation & \textbf{93.9} & 6.1 & 0 & 0 & 0 \\
Max degree & 11 & \textbf{54.4} & 14 & 19.5 & 1.1 \\
K-shell & 9.4 & \textbf{43.6} & 11.9 & 16.2 & 18.9 \\
Betweenness & 13.1 & \textbf{64.5} & 11.7 & 10.6 & 0.1 \\
\hline
\end{tabular}
\label{table_cummunity}
\end{table}

\begin{figure}
\begin{center}
\includegraphics[width=8cm]{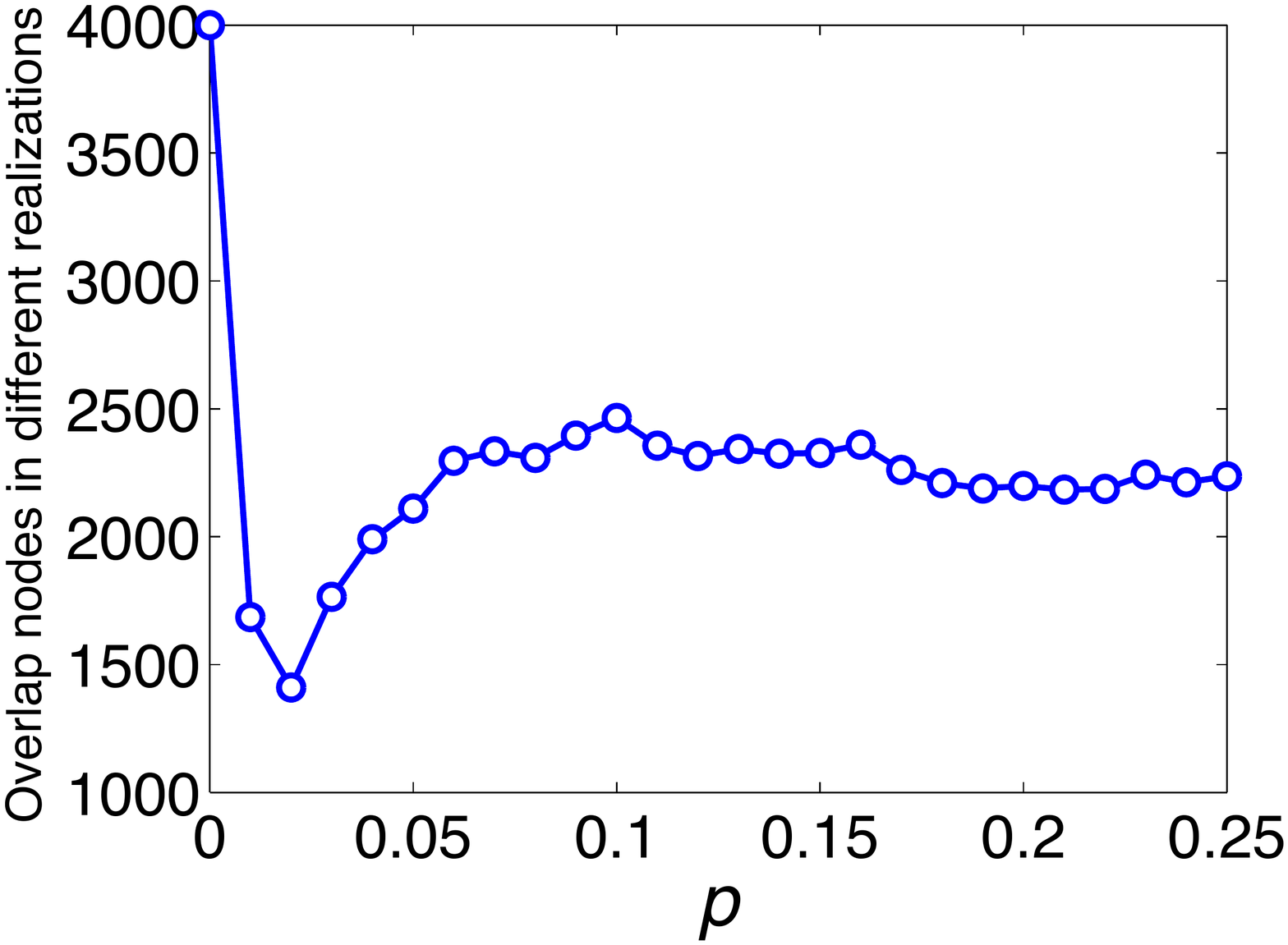}
\includegraphics[width=8cm]{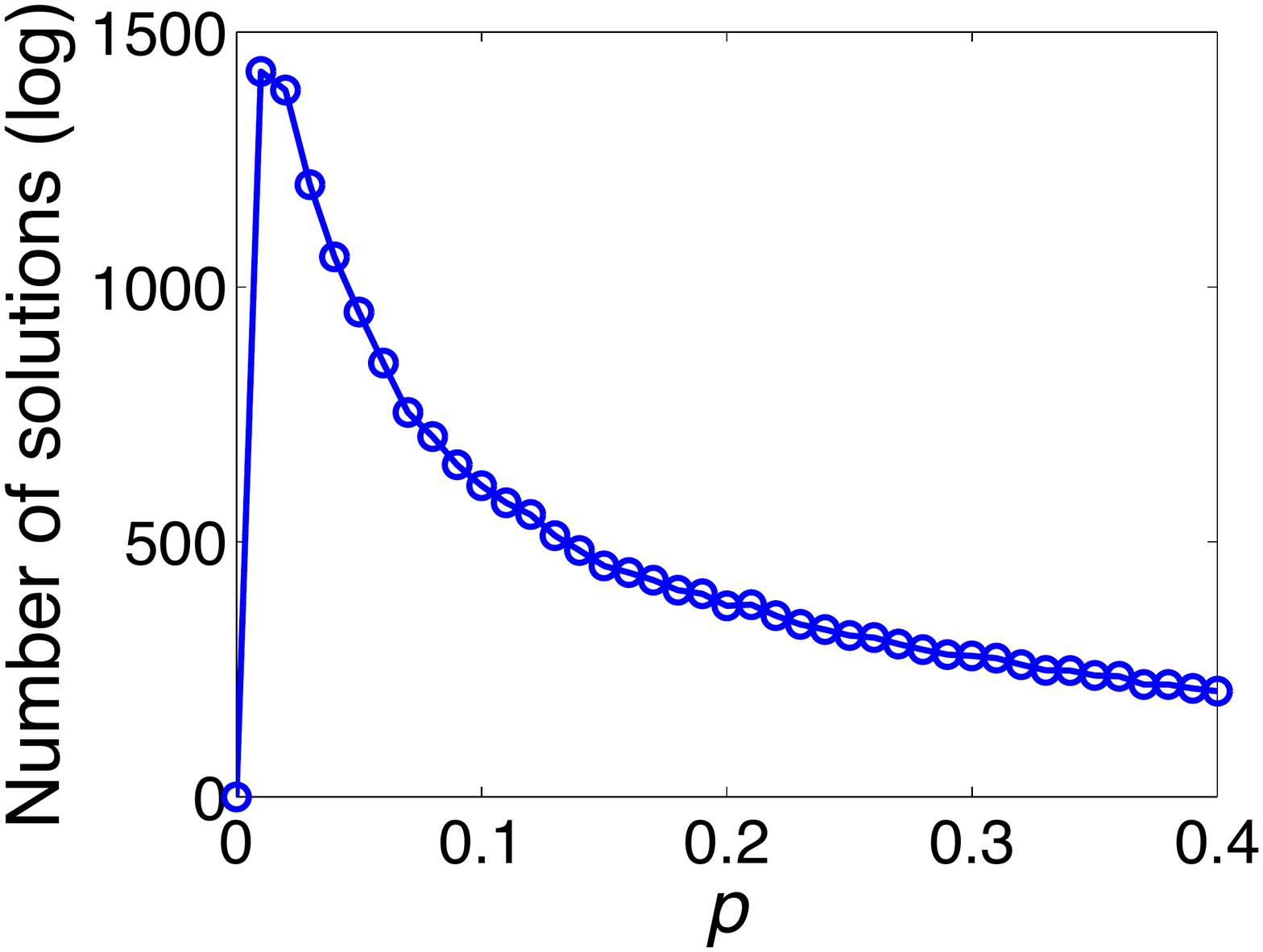}
\mbox{\hspace{1cm} (a) \hspace{7.5cm}  (b)}\\
\caption{The diversity of the spreaders identified by the percolation method on the Facebook network. (a) The number of common nodes between different realizations of the percolation method on the Facebook network. (b) The scale of solution  space. We use a logarithmic scale for better presentation. In each realization, we select 4000 nodes by percolation method. }\label{overlap}
\end{center}
\end{figure}

Most of the other methods always lead to the same set of spreaders. In our percolation method, different set of spreaders may be generated from different realizations, especially for large $p$. Figure~\ref{overlap} (a) shows the number of initial spreaders which are common among different realizations of the percolation method applied on the Facebook network. It is clear that when $p$ increases, the number of common spreaders decreases, indicating that the solutions become more and more diverse. This result has practical significance, especially when some of the initial spreaders are offline, we can use the next best candidate as a back-up spreader without losing spreadability. On the other hand, Fig.~\ref{overlap} (b) shows the entropy of the obtained solution, i.e. the logarithm of the number of different identified spreaders. Compared with the other three methods, percolation method provides a higher flexibility in the choice of spreaders.

\begin{figure}
\begin{center}
\includegraphics[width=14cm]{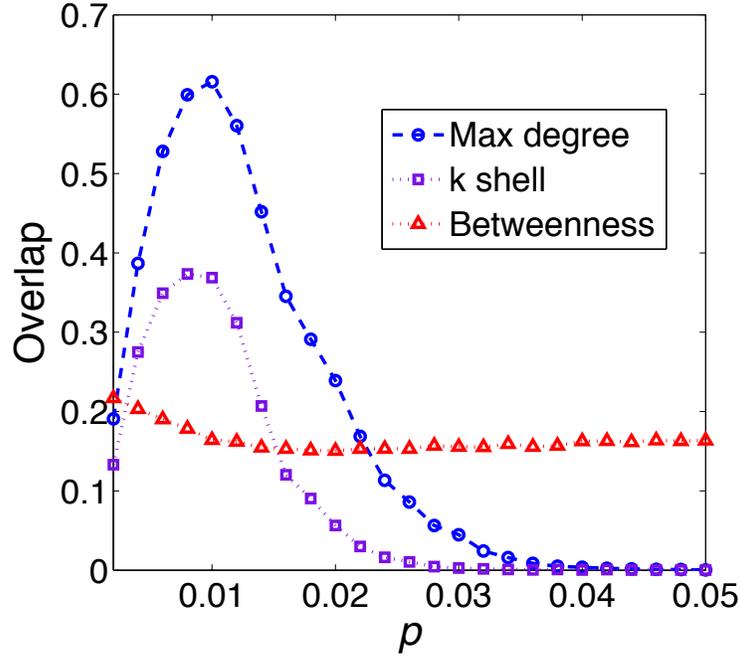}
\caption{The number of common spreaders identified by the percolation method and the other three methods on the Facebook network. We have set $W=L=500$ for the percolation method.}\label{overlap4}
\end{center}
\end{figure}

In order to further examine the difference between our method and the other methods, Fig.~\ref{overlap4} shows the number of identified spreaders which are common between the percolation method and the other methods (comparisons with some other methods are found in Sec. SIII Fig S4 of the SI). The overlap between the percolation method and the degree centrality method reaches the highest value ($\approx 61.5\%$) at the critical point $p_c=0.01$ and then sharply decreases to less than $5\%$ when $p=0.03$. It is because when $p$ increases, most of the high-degree nodes are replaced by nodes with lower degree, and there are a lot of sets of identified spreaders generated from the different realizations as we have discussed in Fig.~\ref{overlap}. What is more impressive is that by increasing the value of $p$ the cost can sharply decrease without losing spreadability and substantially increasing coverage redundancy (see Fig~\ref{example}(c) and ~\ref{example}(e)).

We show in Fig. 5 the relation between the spreadability and the cost. The percolation method is the most cost effective method in terms of spreadability. Four cases are presented, namely $p=0.008<p_c$, $p=0.01=p_c$, $p=0.012$ and $p=0.02>p_c$. Clearly, with the same cost, the percolation method lead to a higher spreadability than the methods of $k$-shell and degree centrality. Although the cost for using betweenness is low, its spreadability is very limited and become saturated at small cost. The percolation method has the highest saturated value of spreadability.

\begin{figure}
\begin{center}
\includegraphics[width=7cm]{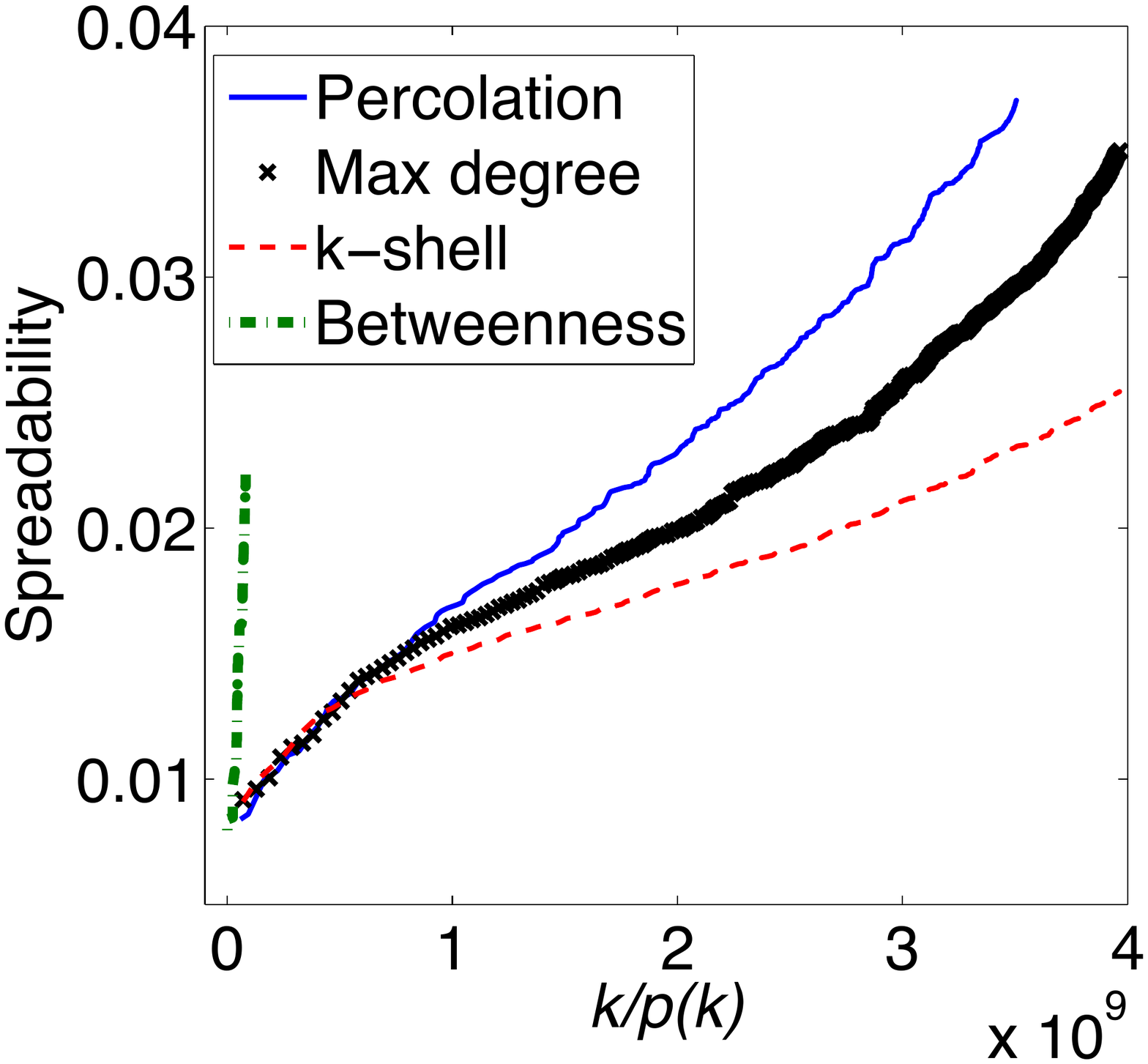}
\includegraphics[width=7cm]{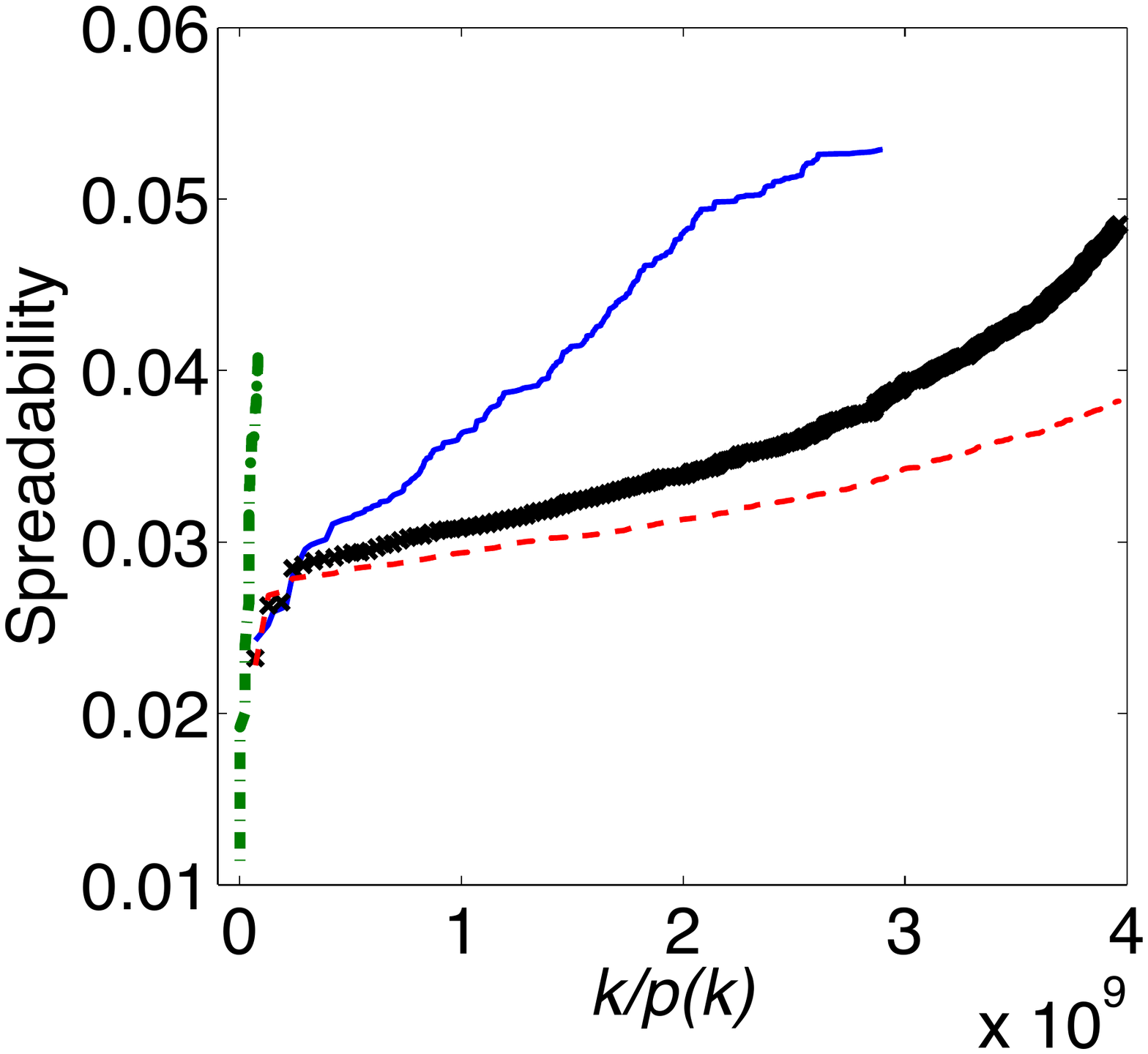}\\
\mbox{\hspace{1cm}(a) $p=0.008$ \hspace{5.5cm}  (b) $p=0.01$}\\
\includegraphics[width=7cm]{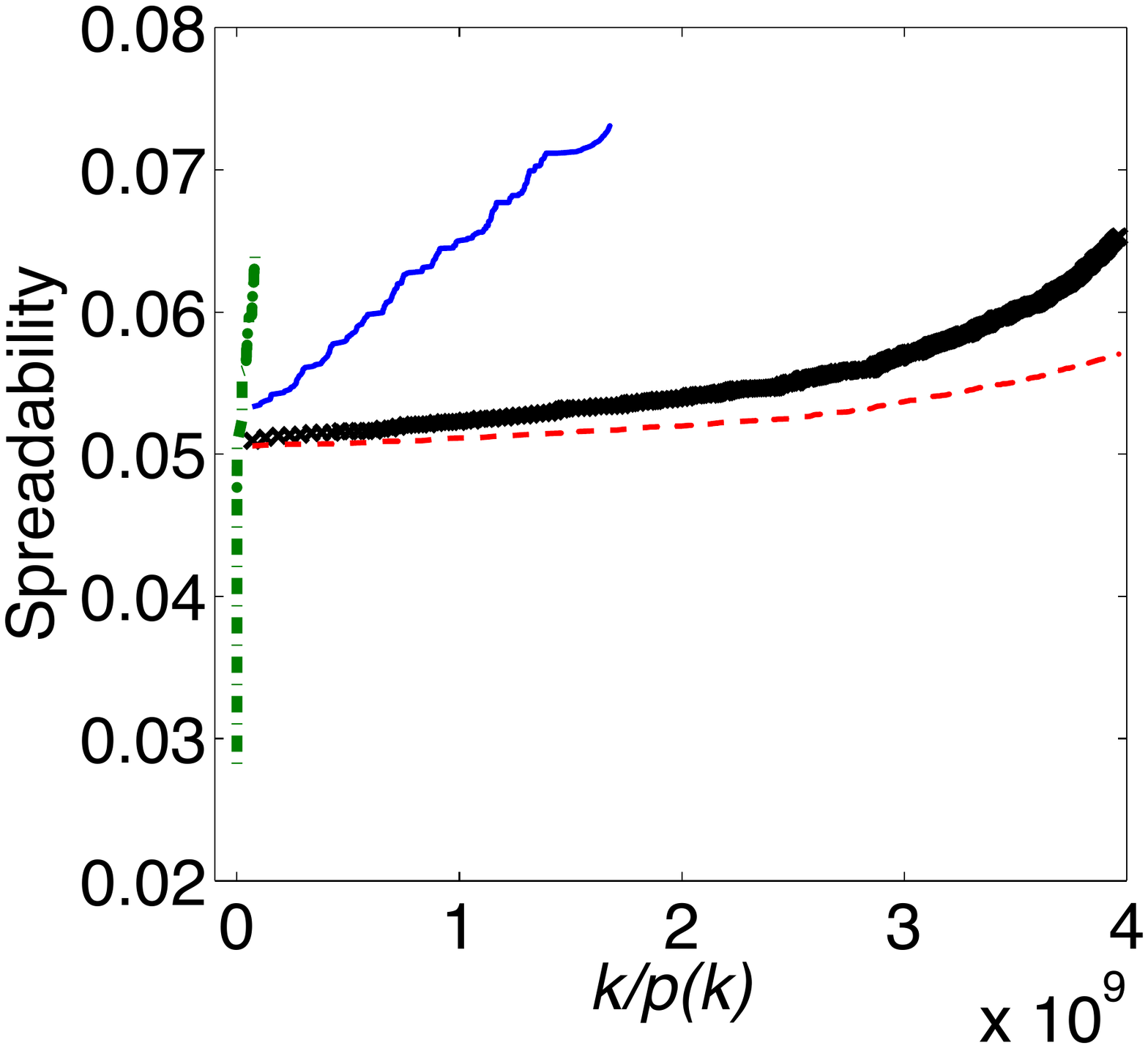}
\includegraphics[width=7cm]{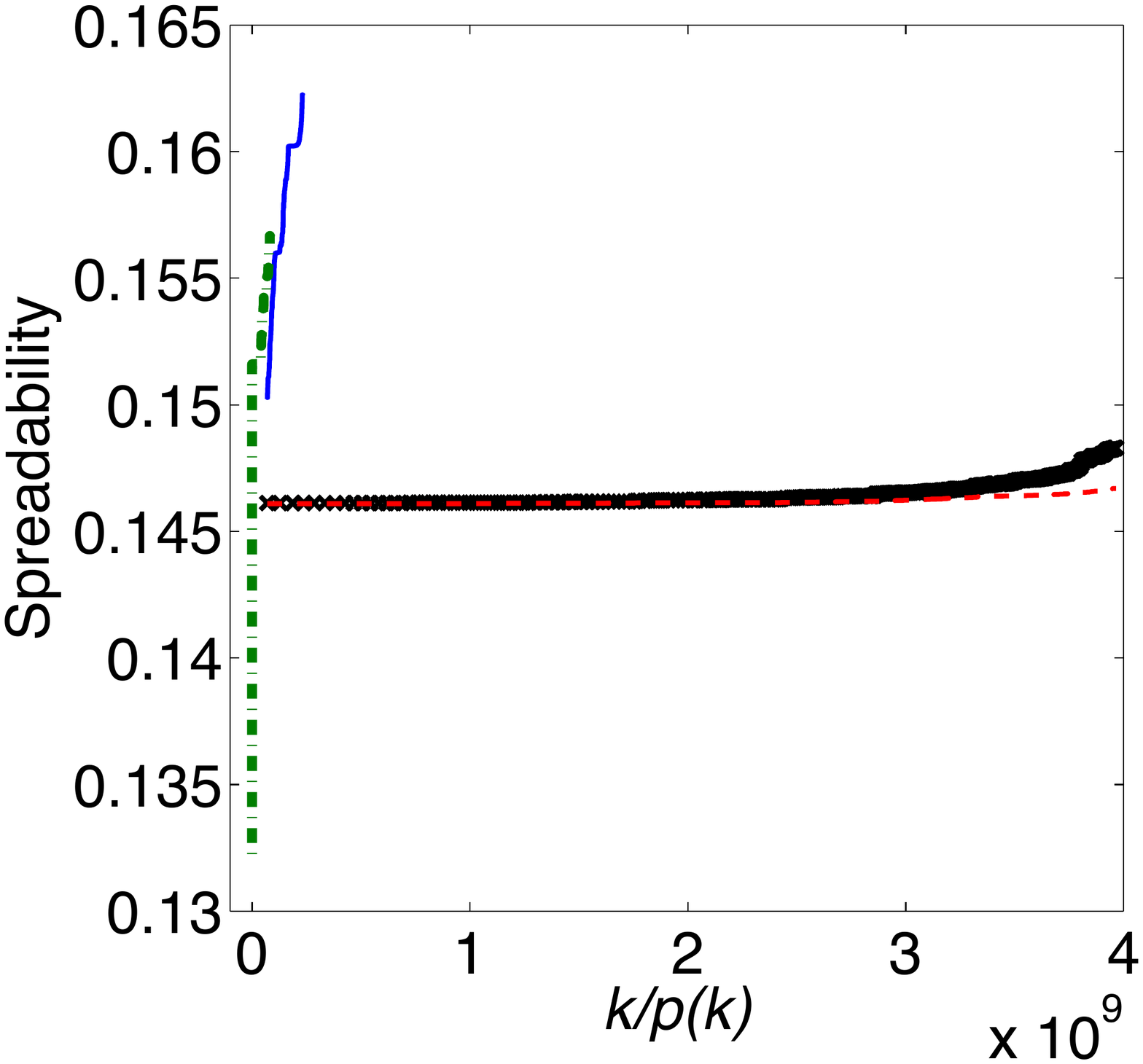}\\
\mbox{\hspace{1cm}(c) $p=0.012$ \hspace{5.5cm}  (d) $p=0.02$}\\
\caption{The spreadability versus the cost to identify spreaders on the Facebook network. We selected 4000 nodes and set $L=4000$ in the percolation method. Here we present a typical case as examples. For a given $p$, we select 4000 top-ranked nodes according to the different methods. For example, to obtain the results of the percolation method (i.e. blue curve), we draw the total cost of the top-$k$ ($k=1,2,3,\cdots,4000$) nodes as $x$-axis value and their corresponding spreadability (as a group) as $y$-axis value. }\label{costp}
\end{center}
\end{figure}

\section{Discussion}

As we can see, social networks constitute a new platform to propagate information. Unlike the usual practice where the networks are used by uncoordinated individuals to share their own message, intended spreading of information can indeed be implemented via the networks. To measure its performance, one can measure the coverage, the redundancy in propagation, and the cost in identifying appropriate initial spreaders. Yet these measures of performance are largely dependent on the choice of users who start the propagation, and there is not a single protocol which achieves optimality in all these dimensions. These difficulties of identifying influential spreaders makes information propagation via social networks remain in its immature stage.

To tackle the challenge, we draw an analogy between the percolation process and information propagation to develop a protocol which gives rise to a low-cost, minimally redundant set of initial spreaders leading to a large coverage. Our protocol was tested on the Facebook network, where favorable results over all the tested centrality-based methods were obtained. When compared to these conventional methods which identify a set of un-coordinated spreaders, the spreaders identified by our protocol are evenly distributed within the network which greatly increases the propagation coverage and reduces its redundancy. Such coordination of spreaders is essential and can only be obtained using the suggested percolation procedures.

The success of this method is not just a coincidence, but it makes the best use of the similarities between percolation and the process of information propagation. By removing edges at random until percolation ceases, we identify individual isolated clusters where news can be effectively propagated within the clusters but not across the clusters. Specific spreaders at the center of these clusters are then identified to be the influential initial spreaders in the original network. By initiating news propagating from this set of spreaders, coverage is increased and redundancy is reduced compared to the conventional centrality methods. Percolation is thus at the center of our propagation protocol instead of a mere analogy.

The remaining question is practicality. As we have discussed, the computational complexity of our protocol is $O(|E|)$, which is a favorable characteristics for applications on practical systems as its complexity scales linearly with the system size. Once the set of important initial spreaders is identified, an coordinator just has to connect to these users and pass the news to them, and information will then propagate quickly throughout the network. Of course, a lot of details and practical difficulties are omitted in this simple description, but our results have lead to insights into a completely new paradigm of information propagation. Further research along this line may revolutionize our way of spreading and gathering information in the near future.

\section{Methods and Materials}
\subsection{Baseline methods}

To identify the most influential spreaders, various centrality measures have been proposed. The first method by which we compare our result with is degree centrality. Degree centrality is a straightforward and efficient metric. It assumes that a node with more nearest neighbors has a higher influence. However, node degree can only reflect its direct influence but not the indirect influence triggered by its nearest neighbors. For example, a node of small degrees, but with a few highly influential neighbors may be more influential than a node having a larger number of less influential neighbors.

The second method we used for comparison is the $k$-shell decomposition. Recent research shows that the location of a node in a network may play a more important role than its degree. A node located in the center of the network is more influential than a node having a larger number of less influential neighbors. Similar to this rationale, Kitsak et al. \cite{Kitsak} proposed a coarse-grained method by using the method of k-core decomposition to quantify the influence of a node based on the assumption that nodes in the same shell have similar influence, and nodes in higher-level shells are likely to infect more nodes.

In the last method, we employ global information to identify the influential spreaders. Specifically, betweenness is one of the most popular geodesic-path-based ranking measures. It is defined as the fraction of shortest paths between all node pairs that pass through the node of interest. Betweenness is, in some sense, a measure of the influence of a node in terms of its role in spreading information \cite{Guimera2002,Yan2006}. For a network $G = (V, E)$ with $n = |V|$ nodes and $m = |E|$ edges, the betweenness centrality of node $v$, denoted by $B(v)$ is \cite{Freeman1977,Freeman1979}
\begin{equation}
B(v)=\Sigma_{s\neq v,s\neq t, v\neq t}\frac{g_{st}(v)}{g_{st}}
\end{equation}
where $g_{st}$ is the number of shortest paths between nodes $s$ and $t$, and $g_{st}(v)$ denotes the number of shortest paths between nodes $s$ and $t$ which pass through node $v$.

\subsection{Computational complexity}
Given a network $G(V,E)$, there are four steps to find the $W$ influential spreaders by the percolation method. Firstly, all the edges are first removed and then recovered with a probability $p$; we then obtain a new network $G'$. The required computational complexity is $O(|E|)$. Secondly, we find the strongly connected components of $G'$ using Tarjan's algorithm \cite{Tarjan} which has a complexity of $O(|V|+|E|)$. Thirdly, we select one node with the highest degree in each of the $L$ largest components and assign one score to the selected nodes. This complexity for the procedures is $O(L*|V|)$. Repeating the above three steps for different realizations, we rank the nodes according to their scores in descending order, and the top-$k$ nodes are chosen to be the most influential spreaders. The different realizations of the percolation process can be computed in parallel and the complexity of each implementation is $O(|E|+|V|+|E|+L\cdot|V|)$. Considering $\langle k\rangle=\frac{2|E|}{|V|}$, then the complexity is $O[(\langle k\rangle+L+1)\cdot|V|]$. Since $\langle k\rangle \ll |V|$  in real networks, then we have $O[(\langle k\rangle+L+1)\cdot|V|]\sim O(|V|)$, i.e. the complexity of our method grows linearly with system size.

\subsection{Model networks with community structures}
There are three steps to generate a network with community structures. In our experiment, we consider a network with $2000$ nodes which has four communities each of which contains 500 nodes. First, we generate a random network of size 500 and with node degrees distributed in power-law with exponent 2.2 using the configuration model  \cite{ catanzaro05}. The minimum degree is 1 and the maximum degree is $\sqrt{500}\approx 23$ \cite{Dorogovtsev2101}. Second, we repeat the above procedures to generate independently the other three networks. Finally, for each pair of sub-networks we randomly selected a fraction $q$ of node pairs to connect them.

\subsection{Datasets}
Datasets we used are described in Sec. I of the SI and the statistical features of the real networks are summarised in Table S1.

\subsection{SIR model and bond percolation}

Susceptible-Infected-Recovered (SIR) model \cite{Anderson1992} is usually used to mimic the spreading processes of disease. Individuals in this model are classified in three states: susceptible ($S$, does not carry the disease and will not infect others but can be infected), infected ($I$, carry the disease and can infect others), recovered ($R$, either dead or recovered from the disease and immune to further infection). The simulation runs in discrete time steps. At each time step, infective individuals transmit the disease to his or her neighbors with probability $\beta$ and will recover with probability $\gamma$. Then the SIR transmissibility is $p=\beta/\gamma$. The process stops when there is no infected node anymore.

The SIR model can be mapped to a bond percolation model where each link exists with a probability equals to the SIR transmissibility $p$ \cite{Newman2002}. After removing the other edges, a number of clusters are formed. It is clear that the ultimate size of the SIR epidemic outbreak is triggered by a single initially infected node, which is precisely the size of the percolation cluster  that the initial node belongs to. Apparently, the nodes in the same cluster are expected to have the same coverage. A review article on epidemic processes in complex networks can be found in Ref. \cite{PastorSatorras}.

\section{Acknowledgement}
This work is partially supported by the NSFC Grant Nos. 61203156, 11205042, and the Fundamental Research Funds for the Central Universities Gran No. 2682014RC17. LL acknowledges research start-up fund of Hangzhou Normal University under Grant PE13002004039 and the EU FP7 Grant 611272 (project GROWTHCOM). CHY acknowledges the Internal Research Grant RG 71/2013-2014R of the Hong Kong Institute of Education.

\bigskip

\bibliographystyle{unsrt}

\end{document}